\documentclass[12pt]{revtex4}
\usepackage{amsfonts}
\usepackage{amsmath}
\usepackage{amssymb}
\usepackage{graphicx}
\usepackage{color}
\usepackage{epsfig}

\usepackage{fancybox}

\newcommand{\Lpf}{\mathcal L^{pf}}
\newcommand{\Lp}{\mathcal L^{p}}
\newcommand{\Lem}{\mathcal L^{em}}

\newcommand{\gradr}{\partial {\bf r}/\partial {\bf R}}

\begin{document}

\title{Noether Theorem of Relativistic-Electromagnetic Ideal Hydrodynamics}
\author{J. H. Gaspar Elsas} \email{jhelsas@gmail.com}
\author{T. Koide} \email{tomoikoide@gmail.com}
\author{T. Kodama} \email{kodama.takeshi@gmail.com}
\affiliation{Instituto de F\'{\i}sica, Universidade Federal do Rio de Janeiro, C.P.
68528, 21941-972, Rio de Janeiro, Brazil}

\begin{abstract}
We present a variational approach for relativistic ideal hydrodynamics interacting with 
electromagnetic fields.
The momentum of fluid is introduced as the canonical conjugate variable of the position of a fluid element,  
which coincides with the conserved quantity derived from the Noether theorem. 
We further show that our formulation can reproduce  
the usual electromagnetic hydrodynamics which is obtained so as to satisfy the conservation of 
the inertia of fluid motion.
\end{abstract}

\maketitle

\section{introduction}

Hydrodynamics has been introduced to describe collective dynamics of many particle systems and widely applied to describe flow phenomena in 
many branch of physics. 
For example, a relativistic hydrodynamic model is applied to analyze collective behaviors of the matter produced in relativistic heavy-ion collisions reproducing various qualitatively aspects of collective dynamics.

Microscopically, the dynamics of the constituent degrees of freedom of such a matter 
is described by Quantum Chromodynamics (QCD), and thus, the symmetry associated with the 
QCD Lagrangian plays essential roles. 
Then the relation between symmetries and hydrodynamics is sometimes a subject to debate. 
For example, there are several proposals to incorporate the effect of the chiral symmetry to relativistic hydrodynamics \cite{kodama}.
One of promising approaches for this purpose  
is to construct a variational approach for hydrodynamics introducing a corresponding symmetry, but 
the applicability of such a variational approach is not trivial because of the following reasons.

One reason is the difficulty of the discussion of dissipation.  
The variational approach of relativistic hydrodynamic is already discussed in Refs. \cite{else}, 
but it is only for the ideal fluid case. 
As a matter of fact, it is known that the traditional variational principle is not applicable 
when there are dissipative terms \cite{koide}.
Another reason is the strong reduction of the number of the degrees of freedom in the derivation of hydrodynamics through a coarse-graining procedure.
Thus it is not clear whether the symmetry expressed in terms of microscopic variables can be still 
re-written by the remaining reduced number of hydrodynamic variables.

Nevertheless, it is worth discussing symmetry of hydrodynamics in terms of the variational approach, 
in particular, in the derivation of a hydrodynamic model coupling with electromagnetic fields.
It is because the momentum conservation is not clear in such a model.
In the phenomenological derivation of hydrodynamics, the conservations of energy and momentum are employed, 
and the latter is normally identified with the conservation of the inertia of fluid (sometimes called kinetic momentum). 
This identification is, however, not reliable when there is the interaction with electromagnetic fields 
because canonical momentum is affected by gauge fields and does not coincides with inertia anymore.
In other words, a model of electromagnetic hydrodynamics is constructed 
without referring to the momentum conservation explicitly \cite{emhydro}.
Then it could be asked whether such a phenomenological argument is still consistent with 
the gauge symmetry and hence the momentum conservation, 
although the derived result in this manner is gauge invariant and seems to be a reasonable generalization 
of the well-known dynamics of charged particles interacting with electromagnetic fields.

The purpose of this work is to derive relativistic-electromagnetic ideal hydrodynamics 
in the variational approach where Lagrange fluid elements are treated as particles.
Then we define the canonical momentum as the conjugate variable of the position of a fluid element.
Applying the variational principle and the Noether theorem to this gauge invariant Lagrangian, 
we can construct the relativistic hydrodynamic model satisfying the momentum conservation in the literature.

As was mentioned, the variational formulation of relativistic ideal hydrodynamic is already discussed 
in Refs. \cite{else} in terms of the Euler coordinates.
However, to discuss the canonical momentum and Noether theorem, 
the formulation in the Lagrange coordinates is much more useful. 
The Noether theorem of the relativistic magneto hydrodynamics is discussed in Ref. \cite{ach}.
However, the derived hydrodynamic equation (given by Eq. (30)) does not coincide 
with the usual hydrodynamic theory. 
Moreover, special conditions associated with the magneto hydrodynamics (given by Eq. (3a)) is employed in the derivation. 
Our result is consistent with the usual relativistic ideal hydrodynamics and both of electric and magnetic fields 
are treat on an equal footing.

This paper is organized as follows.
The Lagrangian density in the Euler coordinates is re-expressed in the Lagrangian coordinates in Sec. 
\ref{sec:2}. In Sec. \ref{sec:3}, we define the canonical momentum of fluid and derive the Euler-Lagrange equation.
In Sec. \ref{sec:4}, the Noether theorem is applied and the energy and momentum conservations are investigated.
Section \ref{sec:5} is devoted to the concluding remarks.

\section{Lagrangian and set up of hydrodynamic variables} \label{sec:2}

A formally gauge invariant Lagrangian of a relativistic ideal fluid is given by  
\begin{equation}
L = \int d^3 {\bf x} \left[ - \varepsilon^* (n^*_e , s^*) - n^*_e u_\mu A^\mu 
- \frac{1}{4} F^{\mu\nu}F_{\mu\nu} \right], \label{lag-euler}
\end{equation}
where 
\begin{equation}
F^{\mu\nu}({\bf x},t) = \partial^\mu A^\nu ({\bf x},t)- \partial^\nu A^\mu({\bf x},t) .
\end{equation}
Here $\varepsilon^*$, $n^*_e$ and $s^*$ are the proper scalar densities which 
coincide with the energy, charge and the entropy densities in the local rest frame, respectively.
The fluid velocity $u^\mu$ is normalized as $u^\mu u_\mu = 1$. The gauge filed is denoted by $A^\mu$.
This Lagrangian is written in the Eulerian coordinates. 
See also Ref. \cite{else}.
Note that there is a equation of continuity associated with the charge conservation, $\partial_\mu (n^*_e u_\mu) =0$. 
By using this property, we can show the gauge invariance of this Lagrangian.

Note that the above Lagrangian is different from the one used in Ref. \cite{ach}.
One can easily see that our Lagrangian is a natural generalization of the one-particle Lagrangian 
with the gauge interaction.

In order to use thermodynamic relations, we assume that $\varepsilon^*$ is given by a function of $n^*_e$ and $s^*$. 
Note that, in the present case, the energy flow can be chosen to be parallel to the conserved charge flow 
and hence $\varepsilon^*$ coincides with the energy density in the same frame where $n^*_e$ agrees with the charge density.
Thus we can employ the following thermodynamic relation,
\begin{equation}
d \varepsilon^* = T ds^* - \mu dn^*_e .
\end{equation}
It should be mentioned that even thermodynamic relations can be modified when fluids interacting with electromagnetic fields in general 
because this interaction contributes the thermodynamic work. 
This contribution, however, vanishes when the polarization and magnetization of the fluid are no considered as is the present case \cite{emhydro}.

As was mentioned, there are two different approaches to describe fluids, one is in the Euler coordinates and the other in the Lagrangian coordinates. The Euler coordinates are fixed in a given referential frame. 
 
\begin{center}
\begin{figure}[t]
\includegraphics[scale=0.4]{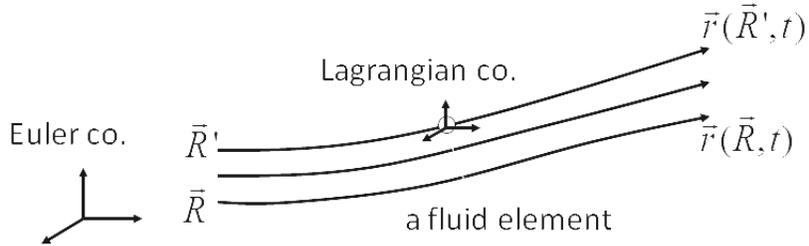}
\caption{The Euler coordinates are fixed in a certain frame. The Lagrangian coordinates are fixed on a certain fluid element.
The trajectory of a fluid element is denoted by ${\bf r}({\bf R},t)$ with ${\bf R}$ being an initial position.}
\label{fig:Eulerco-Lagrangianco}
\end{figure}
\end{center}

The Lagrangian coordinates are, on the other hand, fixed on a certain fluid element and changes its direction with time
as is shown in Fig. \ref{fig:Eulerco-Lagrangianco}. In the present work, we employ the variation in the Lagrangian
coordinates. For this, we introduce a fluid element with a constant entropy $\nu$ 
which is determined by the total entropy divided by the total number of the fluid element. 
Note that the total entropy is conserved for ideal fluids.
Then the trajectory of a fluid element is represented by ${\bf r}({\bf R},t)$ where ${\bf R}$ denotes the initial position.

We further assume that, for any $t$, ${\bf r}({\bf R},t) \neq {\bf r}({\bf R}',t)$ with ${\bf R'}\neq {\bf R}$, and thus this argument is not applicable when turbulence appears. 
Let us denote the volume density of this fluid element by $\rho({\bf r}({\bf R},t),t)$. 
If there is no chaotic motion like turbulence, the volume of each fluid element is conserved and the evolution is determined by 
\begin{equation}
\frac{d}{dt} \rho({\bf r}({\bf R},t),t) = - \rho({\bf r}({\bf R},t),t) (\nabla \cdot {\bf v}({\bf r}({\bf R},t),t) ),  
\label{eq_vol}
\end{equation}
where ${\bf v}({\bf r}({\bf R},t),t) = d{\bf r}({\bf R},t)/dt$.
This is the equation of continuity in the Lagrangian coordinates.
By using this, one can easily see the conservation of the proper scalar density associated with the entropy, 
\begin{equation}
\partial_\mu (s^* u^\mu) = 0,
\end{equation}
where $x^\mu = (ct,x,y,z)$ and $s^* = \nu \rho/\gamma$ with $\gamma = 1/\sqrt{1 - ({\bf v}/c)^2} = u^0$.

\begin{center}
\begin{figure}[t]
\includegraphics[scale=0.2]{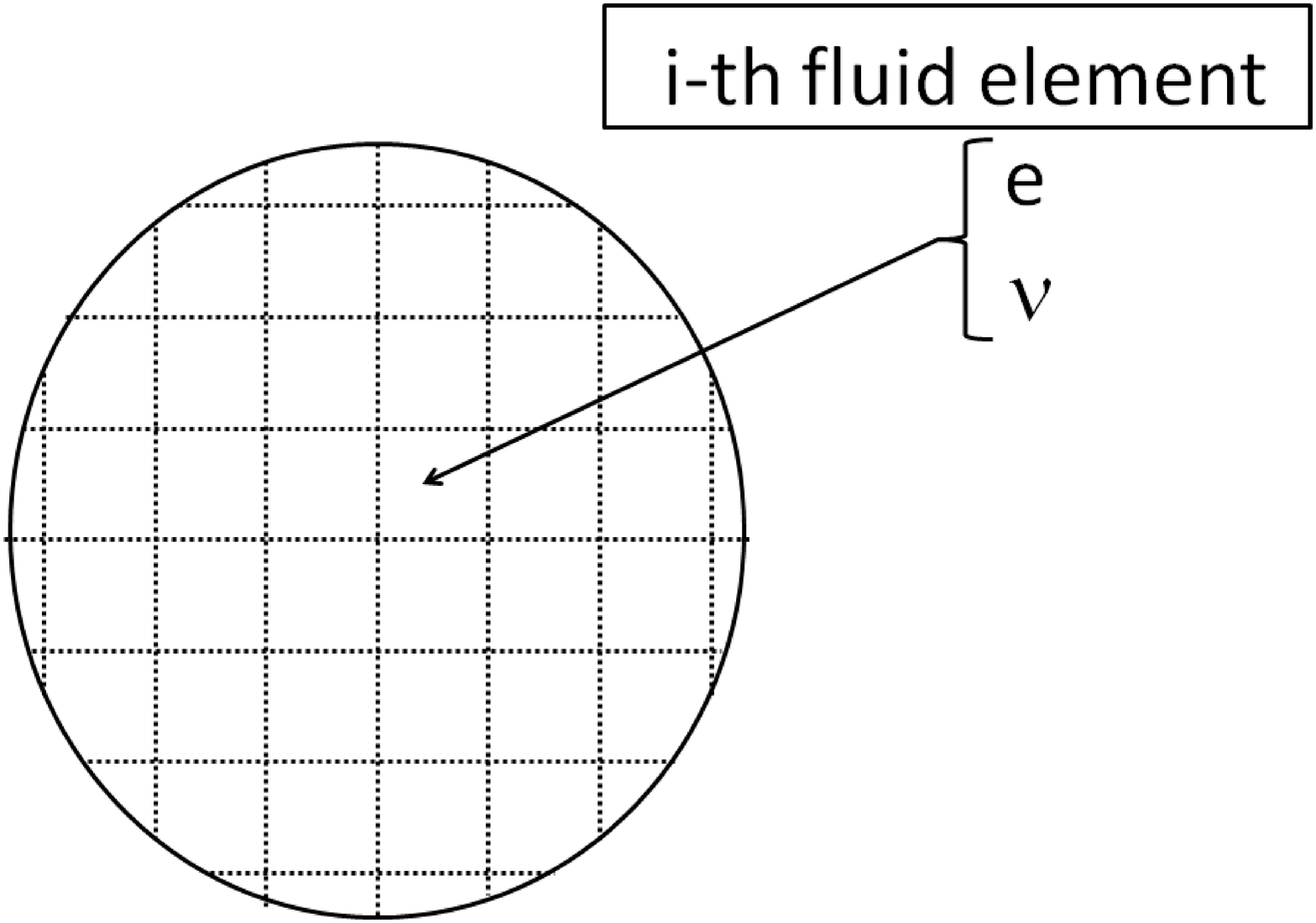}
\caption{Fluid is decomposed into the ensemble of fluid elements. 
The initialy given electric charge and entropy for each fluid element are conserved 
during the hydrodynamic evolution.}
\label{fig:fluidelement}
\end{figure}
\end{center}

In this paper, we consider an ideal fluid, that is, there is no dissipative flow such as viscosity. 
Then the electric charge current also should be given by 
\begin{equation}
\partial_\mu (n^*_e u^\mu) = 0.
\end{equation}
To satisfy this, we need to assume that each fluid element has the same amount of electric charge $e$ which 
is defined by the division of the total electric charge by the total number of the fluid elements, and 
$n^*_e = e \rho/\gamma$. 
In short, as is shown in Fig. \ref{fig:fluidelement}, each fluid element has the same amounts of $\nu$ and $e$.
We will discuss the physical meaning of this assumption in the concluding remarks. 
It is also noted that we cannot argue electromagnetic fields whose typical length scale is smaller than the size of the 
fluid element.
By using this conservation of charge, one can easily prove the gauge invariance of the Lagrangian given by Eq. (\ref{lag-euler}).

If there is no turbulent behavior, the evolution of the Lagrangian coordinates can be regarded as the variable transform from ${\bf R}$ to ${\bf r}$. 
We then find that the Jacobian of the variable transform is expressed as \cite{koide}
\begin{equation}
J({\bf R},t) = {\rm det} \left| \frac{\partial {\bf r}}{\partial {\bf R}} \right|
 = \frac{\rho_0({\bf R})}{\rho({\bf r}({\bf R},t),t)},
\end{equation}
where $\rho_0$ is the initial distribution of the fluid element, $\rho_0({\bf R}) = \rho({\bf r}({\bf R},t_0),t_0)$ 
with $t_0$ being the initial time.
In short, the Lagrangian is obtained by the sum of the contributions from each fluid cells and thus we have 
\begin{eqnarray}
L = \int d^3 {\bf R}  {\cal L}^p ({\bf r},A)  + \int d^3 {\bf x} {\cal L}^{em} (A) ,\label{eq:action}
\end{eqnarray}
where 
\begin{eqnarray}
{\cal L}^p({\bf r},A) 
&=& 
{\cal L}^{pf}({\bf r}) - j^\mu_{e0} A_\mu = - \frac{\rho_0}{\rho} \varepsilon^*(n^*_e,s^*) 
- j^\mu_{e0} A_\mu , \\
{\cal L}^{em} (A) 
&=& -\frac{1}{4} F^{\mu\nu}F_{\mu\nu} ,
\end{eqnarray}
with 
\begin{eqnarray}
j^0_{0} &=& e_{\bf R} \rho_0({\bf R}), \\
j^i_{0} &=& e_{\bf R} \rho_0({\bf R}) \frac{1}{c} v^{i} ({\bf r},t) , \\
F^{\mu\nu} &=& \partial^\mu A^\nu - \partial^\nu A^\mu.
\end{eqnarray}
The gauge field $A^{\mu}$ is introduced to express electromagnetic fields.
It should be emphasized that ${\bf r}$ denotes a trajectory of a fluid element and ${\bf x}$ the spatial coordinates.

\section{Canonical momentum and Euler-Lagrange equation} \label{sec:3}

In the present work, we assume that the canonical momentum of fluid is given by the conjugate variable of the velocity of a fluid element. 
Then we adapt the following definition, 
\begin{equation}
\frac{\partial {\cal L}}{\partial \dot{\bf r}} 
= \rho_0 \frac{\varepsilon^* + P^*}{\rho^*} \gamma \frac{\bf v}{c^2} + \frac{e\rho_0}{c}{\bf A}.
\end{equation}
For the sake of later convenience, however, we exclusively use the canonical momentum expressed in the Euler coordinates, 
\begin{equation}
p^{i} ({\bf x},t) = \frac{\gamma}{c} 
\left\{ (\varepsilon^* + P^*) u^{i} + n^*_e A^{i} \right\},  \label{cano-mom} 
\end{equation}
which satisfies
\begin{equation}
\int d^3 {\bf R} \frac{\partial {\cal L}}{\partial \dot{\bf r}} = \int d^3 x\ {\bf p}({\bf x},t).
\end{equation}

In the non-relativistic limit, the enthalpy density $\varepsilon^* + P^*$ is reduced to $mc^2 \rho^*$ with $m$ being the mass of 
the constituent particle. Then the above definition agrees with the well-known expression 
for the canonical momentum of a non-relativistic particle interacting with electromagnetic fields by 
substituting Dirac's delta function to $\rho^*$.

The variations for the trajectory of a fluid element and the gauge fields are given by 
\begin{eqnarray}
{\bf r}({\bf R},t) &\longrightarrow & {\bf r}'({\bf R},t) = {\bf r}({\bf R},t)+\delta{\bf r}({\bf R},t), \\
A^\nu({\bf x},t)   &\longrightarrow & A^{'\nu}({\bf x},t) = A^\nu({\bf x},t) +\delta A^\nu({\bf x},t),
\end{eqnarray}
respectively.
Then we obtain the following Euler-Lagrange equations, 
\begin{subequations}
\begin{eqnarray}
u^\mu \partial_\mu \left[\frac{\varepsilon^* + P^*}{\rho^*}{\bf u}\right]&=& -\frac{1}{\rho}\nabla_{{\bf r}} P^*+e\left[{\bf E}+\frac{{\bf v}}{c} \times {\bf B}\right], \\
\partial_\mu F^{\mu\nu} &=& n_e^* u^\nu.
\end{eqnarray}
\label{el}
\end{subequations}
Here the thermodynamic pressure is defined by 
\begin{equation}
P^* = Ts^* - \varepsilon^* + \mu n^*_e.
\end{equation}
See also the calculations in Ref. \cite{koide}.

The latter equation is Maxwell's equation.
The former in the Euler coordinates can be expressed in a more familiar form as,
\begin{equation}
\left( \partial_t + {\bf v}\cdot \nabla \right)\left[\frac{\varepsilon^* + P^*}{\rho^*}{\bf u}\right]= -\frac{1}{\rho}\nabla P^*+e\left[{\bf E}+{\bf v} \times {\bf B}\right] .
\end{equation}
This is the well-known momentum component of the relativistic ideal hydrodynamics with the Lorentz force.

\section{Noether theorem} \label{sec:4}

A hydrodynamic model is constructed so as to satisfy the energy-momentum conservation, which 
is obtained as the result of the invariance of actions for the time and space translations, respectively, 
as is well-known in the Noether theorem.

Let us consider the following transform of the coordinate variables, 
\begin{equation}
x^\mu \longrightarrow (x')^\mu = x^\mu + \epsilon^\mu.
\end{equation}
Then the gauge fields and the fluid trajectory are transformed as 
\begin{eqnarray}
A^\mu (x) &\longrightarrow& 
(A')^\mu (x') = A^\mu(x) + \delta^L A^\mu(x) + \epsilon^\nu \partial_\nu A^\mu , \\
{\bf r}({\bf R},t) &\longrightarrow& 
{\bf r}'({\bf R},t') = {\bf r}({\bf R},t) + \delta^L {\bf r}({\bf R},t)+ \epsilon^0 \frac{d}{d(ct)}{\bf r}({\bf R},t). \label{tra-tra}
\end{eqnarray} 
Here the Lee derivative is defined by 
\begin{equation}
\delta^L f^{i}(x) = (f')^{i}(x) - f^{i}(x).
\end{equation}
In the interaction part of the Lagrangian, we need to estimate the transform of the gauge fields on the trajectory of a fluid element 
as  
\begin{equation}
A^\mu ({\bf r},t) \longrightarrow (A')^\mu ({\bf r}',t') 
= 
A^\mu ({\bf r},t) + \delta^L A^\mu 
+ \epsilon^0 \partial_0 A^\mu 
+ \left( \delta^L {\bf r}
+ \epsilon^0 \frac{d{\bf r}}{d(ct)} \right) \cdot \nabla  A^\mu .
\end{equation}

Note that because of the transform of the trajectory of a fluid element (\ref{tra-tra}), 
we cannot implement calculations keeping the Lorentz covariance manifestly, 
but the derived results can be expressed in the manifestly covariant forms.

Substituting these transforms to the action, we obtain 
\begin{eqnarray}
\lefteqn{I' - I} && \nonumber \\
&=& \int dt d^3{\bf R} \left\{\frac{d}{dt}\left[\frac{\epsilon^0}{c} \Lp + \left(\frac{\partial \Lpf}{\partial \dot{\bf r}}+\frac{e\rho_0}{c} {\bf A}\right)\cdot \delta^L {\bf r}\right] + \frac{\partial}{\partial {\bf R}}\cdot \left[\frac{\partial \Lpf}{\partial (\gradr)}\delta^L{\bf r}\right]\right\} \nonumber\\
&& + \frac{1}{c}\int d^4 x\ \partial_\mu \left[\epsilon^\mu \Lem + \frac{\partial \Lem}{\partial (\partial_\mu A^\nu)}\delta^L A^\nu\right],
\label{noether}
\end{eqnarray}
where $I$ and $I'$ denote the action before and after the transform, respectively.
We used the Euler-Lagrange equations (\ref{el}) in the derivation of this expression. 
In the following, we discuss the momentum and energy conservations, separately.

\subsection{Momentum conservation}

The momentum conservation is observed by employing the invariance of the action for the spatial translation, which is expressed as
\begin{eqnarray}
\epsilon^0 &=& 0, \\
\delta^L {\bf r} &=& \vec{\epsilon}, \\
\delta t &=& 0, \\
\delta^L A^\mu &=& -(\vec{\epsilon} \cdot \nabla) A^\mu .
\end{eqnarray}
By substituting these into Eq. (\ref{noether}) and re-expressing in the Euler coordinates, 
we have 
\begin{equation}
\partial_\mu \left[ 
 (T_{fluid})^{\mu i} + (T_{em})^{\mu i}
\right] = 0, \label{eos-mom}
\end{equation}
where 
\begin{eqnarray}
T^{\mu i}_{fluid} &=& (\varepsilon^*+P^*)u^\mu u^i - P^* g^{\mu i} + \frac{n^*_e}{c} u^\mu A^i,  \\
T^{\mu i}_{em} &=& \frac{g^{\mu i}}{2}\partial_\alpha A_\nu(\partial^\alpha A^\nu - \partial^\nu A^\alpha) - \partial^i A^\nu (\partial^\mu A_\nu - \partial_\nu A^\mu) .\label{tmunu-em}
\end{eqnarray}

By using the definition given by Eq. (\ref{cano-mom}), the time component $(T_{fluid})^{0 i}$ is expressed as 
\begin{equation}
(T_{fluid})^{0 i} = c p^{i}. \label{tfluid0i}
\end{equation}
That is, Eq. (\ref{eos-mom}) represents the conservation of the canonical momentum which is defined by the Legendre 
transform (\ref{cano-mom}). In other words, the definition of the canonical momentum of fluid introduced 
in this paper is consistent with the Noether theorem, 
and the hydrodynamics constructed in our formulation conserves the canonical momentum appropriately.

On the other hand, the momentum of electromagnetic fields in the present case, 
from the definition (\ref{tmunu-em}), is given by
\begin{equation}
(T_{em})^{0i} = ({\bf E} \times {\bf B})^{i} - \frac{1}{c}\gamma n^*_e A^{i} -\nabla \cdot(A^{i} {\bf E}). \label{tem0i}
\end{equation} 
The first term on the right hand side is the well-known momentum density of electromagnetic fields, while 
the second term represents the modification by the interaction with the fluid. 
And the corresponding stress tensor is calculated to be
\begin{equation}
(T_{em})^{ji} = -E^j E^i - B^j B^i + \frac{1}{2} \left({\bf E}^2+{\bf B}^2\right)\delta^{ji} - \frac{n^*_e}{c} u^j A^i - \sum_{\alpha=0}^3 \partial_\alpha (F^{j \alpha}A^i).
\end{equation}
Note that the term $\nabla \cdot(A^{i} {\bf E})$ appearing in Eq. (\ref{tem0i}) 
is cancelled by the corresponding term in $\partial_\alpha (F^{j \alpha}A^i)$ when these are substituted into 
the equation of continuity.

The momenta of fluid and electromagnetic fields are modified by interaction as is shown by Eqs. (\ref{tfluid0i}) 
and (\ref{tem0i}). Note, however, that the modifications have the same form but the opposite sign.
In short, the sum of the two momenta is reduced to that of the inertia of the fluid 
and the free electromagnetic momentum density as
\begin{equation}
(T_{fluid})^{0i} + (T_{em})^{0i}
= \gamma (\varepsilon^* + P^*)u^{i} + ({\bf E} \times {\bf B})^{i}.
\end{equation}
Therefore, the conservation of momentum is equivalent to that of inertia as is 
expected in usual phenomenological argument.

\subsection{Energy conservation}

The energy conservation is discussed in a similar fashion.
The time translation is expressed by 
\begin{eqnarray}
\epsilon^i &=& 0, \\
\delta^L {\bf r} &=& - \frac{\epsilon^0}{c} \dot{\bf r}, \\
\delta t &=& \frac{\epsilon^0}{c}, \\
\delta^L A^\mu &=& - \epsilon^0 \partial_0 A^\mu .
\end{eqnarray}
Then Eq. (\ref{noether}) is reduced to
\begin{equation}
\frac{1}{c}\frac{\partial}{\partial t}\left(T^{0 0}_{fluid} +T^{00}_{em}\right) 
+ \sum_i \partial_i \left(T^{i 0}_{fluid} + T^{i0}_{em}\right) = 0,
\end{equation}
where
\begin{eqnarray}
T^{\mu 0}_{fluid} &=& (\varepsilon^*+P^*)u^\mu u^0 - P^* g^{\mu 0} + \frac{n^*_e}{c} u^\mu A^0,  \\
T^{00}_{em} &=& \frac{1}{2}({\bf E}^2+{\bf B}^2) - \frac{1}{c}\gamma n^*_e  A^0 - \nabla \cdot (A^0{\bf E} ), \label{tem00} \\
T^{i0}_{em} &=& ({\bf E}\times {\bf B})^i- n_e^* A^0 u^i - \sum_{\alpha=0}^3 \partial_\alpha (A^0 F^{i\alpha}). \label{temi0}
\end{eqnarray}

We observe that the energy density of the fluid is given by the sum of the non-interacting ideal fluid part
$(\varepsilon^*+P^*)\gamma^2-P^*$ and the contribution from the 
Coulomb energy $\gamma n^*_e A^0$. The same Coulomb contribution appears in the energy density of
electromagnetic fields with the opposite sign.
The third term in Eq. (\ref{tem00}) cancels with a term in Eq. (\ref{temi0}) in substituting into the equation of continuity as is the case of the momentum conservation.
In short, the total energy density which is conserved is expressed as
\begin{equation}
T^{00}_{fluid} + T^{00}_{em} = (\varepsilon^*+P^*)\gamma^2-P^* + \frac{1}{2}({\bf E}^2+{\bf B}^2). 
\end{equation}
Again, the sum of the two energy densities is 
reduced to that of the non-interacting ideal fluid and electromagnetic fields.

In short, these two conservations can be cast into the traditional covariant forms as
\begin{eqnarray}
\partial_\mu (T^{\mu\nu}_{ideal} + T^{\mu\nu}_E) &=& 0,\\
\partial_\mu N^\mu &=& 0, 
\end{eqnarray}
where
\begin{eqnarray}
T^{\mu\nu}_{ideal} &=& (\varepsilon^* + P^*)u^{\mu} u^{\nu} - P^* g^{\mu\nu}, \\
T^{\mu\nu}_E &=& F^{\mu\lambda} F^\nu_{\lambda} + \frac{1}{4}g^{\mu\nu}F^{\lambda\rho}F_{\lambda \rho} ,\\
N^\mu &=& n^*_E u^\mu .
\end{eqnarray}
These are well-known phenomenological results of relativistic-electromagnetic ideal hydrodynamics.

\section{Concluding Remarks} \label{sec:5}

In this paper, we reformulated relativistic-electromagnetic ideal hydrodynamics in the framework of 
the variational approach.
Then we interpreted the canonical momentum of fluid as the canonical conjugate variable of the position of a fluid element.
This canonical momentum is affected by the interaction with gauge fields as is well-known in 
the case of charged particle systems. 
To confirm our assumption for the definition of the canonical momentum of fluid, 
the conservation law associated with the the spatial translational invariance 
of the action is derived by applying the Noether theorem and it is confirmed that this Noether charge is equivalent to the above canonical momentum.
Such a modification of momentum by the gauge interaction appears even for the electromagnetic momentum density. 
However, the sum of the two momenta is finally given by that of fluid inertia and free electromagnetic fields.

This is true even for the definitions of energy densities. The fluid energy density defined by the time translation invariance contains 
the contribution from the Coulomb energy. This contribution is however canceled with the corresponding term in the electromagnetic energy density and 
hence the total energy density is finally given by the sum of the non-interacting ideal fluid and electromagnetic fields.

In the end, our formulation reproduces the usual phenomenologically derived electromagnetic hydrodynamics.
This result suggests us the possibility that 
the symmetry of fluid is possible to be investigated in terms of 
the Lagrangian approach.
The potential application is the modeling of hydrodynamics including the effect of the chiral symmetry, 
which has been discussed by several groups \cite{kodama}.
For this, the chiral symmetry should be re-expressed in terms of hydrodynamic variables.
To see the relation between this symmetry and hydrodynamic variables, 
it will be useful to express the Dirac equation with hydrodynamic variable \cite{dirac}.

In this derivation, we assumed the existence of fluid elements which have a constant entropy $\nu$ and charge $e$ and these 
are simultaneously homogeneous with respect to the Lagrangian coordinates $\partial \nu /\partial {\bf R} = \partial e /\partial {\bf R} = 0$. This is a strong assumption and will be
satisfied only in an idealized situation, because, to satisfy such a condition, the flows of particles and anti-particles
should be, for example, always parallel. This is not likely in realistic hydrodynamic evolution. Thus the argument developed here seems to be 
valid for the case where the chemical potential of a conserved charge is extremely large and the contribution from 
anti-particles are negligibly small compared to that of particles. The same idealization is used for 
the kinetic derivation in Ref. \cite{matsui}.

In the present argument, we have ignored the effect of dissipations, which is necessary to describe more realistic fluids. 
Usually, such an effect is taken into account by introducing the so-called Rayleigh dissipation function. 
In this case, however, the Noether theorem associated with the invariance of actions cannot be obtained. 
Recently, it was shown that the Navier-Stokes-Fourier Equation can be formulated in the framework 
of the stochastic variational method \cite{koide} and the stochastic generalization of the Noether theorem is still applicable in this formulation. 
Thus it is interesting to investigate whether this method is still useful to derive relativistic dissipative hydrodynamics.

\section*{Acknowledgement}

The authors acknowledge financial supports by CNPq, FAPERJ and PRONEX.

\end{document}